\documentstyle[pre,eqsecnum,aps,graphics]{revtex}
\RequirePackage{ifthen}[1994/05/27]
\newboolean{drafton}    \setboolean{drafton}{false}
\ifthenelse{\boolean{drafton}}%
{    \newcommand{\equn}[1]{
        \mbox{}~\hspace*{\stretch{1}}~\begin{picture}(0,0)
           \setlength{\fboxsep}{.7mm}
           \put(3,-12){\makebox[0mm][l]{\fbox{\small #1}}}
        \end{picture}
        \begin{equation}\label{#1}}
    \newcommand{\eqan}[1]{
        \setlength{\fboxsep}{.7mm}
        \mbox{}~\hspace*{\stretch{1}}~\begin{picture}(0,0)
        \put(3,-12){\makebox[0mm][l]{\fbox{\small #1}}}
        \end{picture}
        \begin{eqnarray}\label{#1}}}
{   \newcommand{\equn}[1]{\begin{equation}\label{#1}}
    \newcommand{\eqan}[1]{\begin{eqnarray}\label{#1}}}
\newcommand{\eqa}{\begin{eqnarray}}
\newcommand{\equ}{\begin{equation}}
\newcommand{\nuqe}{\end{equation}}
\newcommand{\uqe}{\end{equation}}
\newcommand{\naqe}{\end{eqnarray}}
\newcommand{\aqe}{\end{eqnarray}}
\newcommand{\nonu}{\nonumber}

\newcommand{\goto}{\rightarrow}
\newcommand{\half}{\frac{1}{2}}

\newcommand{\n}{{\bf \nabla}}

\begin{document}
\twocolumn[\hsize\textwidth\columnwidth\hsize\csname@twocolumnfalse%
\endcsname
\draft

\title{Contact angles on heterogeneous surfaces: \\
a new look at Cassie's and Wenzel's laws}
\author{Peter S.\ Swain \cite{email} and Reinhard Lipowsky \cite{email2}}
\address{Max-Planck Institut f\"{u}r Kolloid- und Grenzfl\"{a}chenforschung,\\
Kantstrasse 55, D-14513 Teltow-Seehof, Germany}
\maketitle

\begin{abstract}
We consider a three dimensional liquid drop sitting on a rough and chemically heterogeneous substrate. Using a novel minimization technique on the free energy of this system, a generalized Young's equation for the contact angle is found. In certain limits, the Cassie and Wenzel laws, and a new equivalent rule, applicable in general, are derived. We also propose an equation in the same spirit as these results but valid on a more `microscopic' level. Throughout we work under the presence of gravity and keep account of line tension terms. 
\end{abstract}

\narrowtext
\vskip2pc]

\section{Introduction}
Recently, there has been an upsurge of interest in wetting phenomena, particularly in the experimental arena with many dramatic new techniques and results available. The impact of these on theory has been twofold, both allowing older work to be thoroughly tested and also posing fresh challenges for those researching in the field today (for current reviews see \cite{rev}). Presently, most theoretical approaches are trying to go beyond wetting on simple, flat, homogeneous substrates and are exploring the effect a structured surface can have on the wetting phase. This heterogeneity can be both geometrical and chemical. However, there have existed for quite some time a few empirical laws describing the contact angle of a drop sitting on such a heterogeneous surface. Wenzel \cite{wenzel} introduced an `average' contact angle on a rough, chemically homogeneous substrate which is expressed in terms of the contact angle on a planar one. Likewise, for smooth but chemically heterogeneous surfaces the Cassie equation \cite{cassie} is widely used, which defines an `average' contact angle by the weighted mean of the angles that the drop would take on pure substrates. The associated wall tensions have been recently studied both for the geometrically rough \cite{borgs,us} and for the chemically heterogeneous case \cite{urban}. Much of the literature has also been concerned with the modification of these laws when line tension effects are included (see, for example, \cite{tosh,drelich,drelich2}). 

In this paper we investigate the statistical mechanical foundations of such empirical approaches. Young's equation \cite{young} is surely the bedrock of all wetting theory and from its most general form, after making an important number of assumptions, we are able to provide a systematic derivation of both the Cassie and Wenzel laws. We choose to describe the chemical heterogeneities within the surface in terms of interfacial and contact line tensions which are position dependent. We consider a solid substrate composed of several different atomic (or molecular) species and first coarse-grain up to a certain length scale, the small scale cut-off, in order to define appropriate composition variables. For the simplest case of a binary system, only one such composition variable is needed and can be defined, for example, as the relative area fraction of one of the two species in the surface layer. In this way, we arrive at composition variables, $X({\bf y})$ say. These, in general, depend on the coordinate ${\bf y} \equiv (y_1,y_2)$ of the two dimensional surface and thus reflect the chemical heterogeneities which are present on length scales large compared to the atomic scale.

A chemically {\it homogeneous} surface is characterised by position independent composition variables $X({\bf y}) = X$ and one may define the different interfacial tensions $\sigma$ and the line tension $\lambda$ in the usual way \cite{rowlinson}. The values of these tensions will, of course, depend on the values of the composition variables: $\sigma = \sigma(X)$ and $\lambda = \lambda(X)$. Thus, in the {\it heterogeneous} case with $X = X({\bf y})$, one may allude to a small gradient expansion and assume that the local tensions are given by $\sigma = \sigma(X({\bf y}))$ and $\lambda = \lambda(X({\bf y}))$, i.e.\ they are essentially determined by the local surface composition. In general, the anisotropy of the solid substrate will lead to a line tension which depends on the orientation of the contact line. In the following, this anisotropy will be ignored and the surface of the solid will be treated as a structureless wall.

For imprinted surfaces, one has surface domains which are large compared to the small scale cut-off \cite{peter}. An example is provided by domains obtained from microcontact printing which have typical sizes in the micrometre range. In this case, the various tensions are constant within the domains but vary across their boundaries. Droplets on such domains may exhibit contact angles which do {\it not} satisfy Young's equation in the limiting case where the boundary width is small compared to the domain size. However, in the absence of such an extreme separation of length scales, one has a position dependent Young equation as discussed in \cite{peter} (though here contributions from the line tension have been ignored).

The outline of our paper is as follows. In Sec.\ \ref{wetting} we specify the free energy of a drop adsorbed on a heterogeneous substrate. By applying a novel minimization procedure to this free energy, detailed in Sec.\ \ref{min}, we are able to find the most general form of Young's equation. A statistical mechanical interpretation of Cassie's and Wenzel's laws is then discussed in Sec.\ \ref{spatial} and a number of strong assumptions highlighted which are implicity adopted by them. In the following sections, using the generalized Young equation, we are then able to go on and provide a systematic derivation of both these rules where we keep account of line tension and gravity. For substrates which have both geometrical and chemical structure a new composite equation is suggested in Sec.\ \ref{both}. An alternative approach is adopted in Sec.\ \ref{alternative} in which a local version of Cassie's law is proposed which needs no suppositions about the drop or heterogeneity of the surface. Finally, we make some short conclusions. 

\section{Wetting of a heterogeneous surface}
\label{wetting}
\begin{figure}[h]
\begin{center}
\scalebox{0.5}{\includegraphics{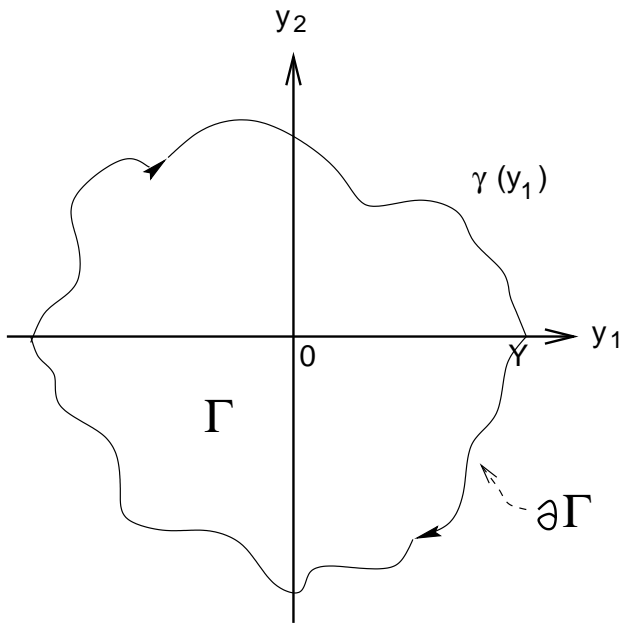}}
\end{center}
\caption{Schematic of the three phase contact line of a droplet sitting on a rough, heterogeneous substrate. The contact area is denoted by $\Gamma$ and the position of the contact line by $\partial \Gamma$. The upper right-hand quadrant defined by $0 \le y_1 \le Y$ is described by $y_2 = \gamma(y_1)$ for some function $\gamma(x)$.}
\label{quad}
\end{figure}

We start by considering the free energy of a three dimensional drop of non-volatile $\beta$ phase within a bulk $\alpha$ phase sitting on a rough, chemically heterogeneous substrate or wall. Let the area covered by the drop be $\Gamma$ and let the edge of this area, i.e.\ the position of the contact line, be $\partial \Gamma$ (see Fig.\ \ref{quad}). 
The $\alpha \beta$ interface has a surface tension $\sigma$ and is at height $h({\bf y})$ above some reference plane, with Cartesian coordinates $(y_1,y_2) = {\bf y}$ as in Fig. \ref{quad}. The line tension of the contact or triple phase line, $\partial \Gamma$, is denoted $\lambda$, while the wall-$\alpha$ and wall-$\beta$ interfacial free energies are $\sigma_{w\alpha}$ and $\sigma_{w\beta}$, respectively. If the substrate surface has a configuration $Z({\bf y})$ above the reference plane then the free energy is 
\eqan{f}
F[h,Z] &=& \int_\Gamma d{\bf y} \Biggl \{ \sigma \sqrt{1+(\n h)^2} - \Bigl[ \sigma_{w\alpha}({\bf y}) - \sigma_{w\beta}({\bf y}) \Bigr] \nonu\\
& & \times \sqrt{1+(\n Z)^2} + {\cal Q}(h,Z) \Biggr \} \nonu \\
& & + \oint_{\partial \Gamma(Z)} ds \; \lambda(s)
\naqe
Here $s$ is the arc length of the contact line at $\partial \Gamma$ on the substrate surface $z=Z({\bf y})$ and $\n$ is the two-dimensional gradient operator, $\n = \left( \frac{\partial}{\partial y_1}, \frac{\partial}{\partial y_2} \right)$. We choose the location of the ${\bf y}$-plane to be such that
\equn{plane}
\int d{\bf y} Z({\bf y}) = 0
\nuqe
The function ${\cal Q}(h,Z)$ in (\ref{f}) can be decomposed into
\equn{P}
{\cal Q}(h,Z) = \Delta p (h-Z) + \half \Delta \rho g (h^2-Z^2)
\nuqe
where $\Delta p$ and $ \Delta \rho$ are the pressure and density differences between the $\beta$ and $\alpha$ phases, respectively. The gravitational acceleration is denoted $g$. Intermolecular forces, such as van der Waals interactions, enter only implicitly through the various tensions \cite{vdw}.  

\section{Minimization procedure}
\label{min}
The equilibrium configuration of the drop will be given by that location $h({\bf y})$ of the liquid surface or $\alpha \beta$ interface and that contact line configuration $\partial \Gamma(Z)$ which minimizes (\ref{f}). To carry out this functional minimization it proves useful to take advantage of a version of Gauss's theorem in two dimensions. For future reference we state this here
\equn{gauss}
\int_\Gamma d{\bf y} (\n f) . {\bf g} = \oint_{\partial \Gamma} f (g_2 dy_1 - g_1 dy_2) - \int_\Gamma d{\bf y} f \n.{\bf g}
\nuqe
where the line integral is evaluated in a clockwise direction as indicated in Fig.\ \ref{quad}.

Rather than functionally minimize with respect to $h({\bf y})$ with the variable boundary condition
\equn{bc}
h({\bf y}) = Z({\bf y}) \mbox{\hspace*{1cm} for ${\bf y} \in \partial \Gamma$}
\nuqe
we choose instead a different approach. Initially $\partial \Gamma$ is fixed, the equilibrium liquid surface calculated for this particular configuration and then the free energy minimized again with respect to the location of the contact line $\partial \Gamma$. 

Using (\ref{gauss}) minimization of (\ref{f}) with respect to $h({\bf y})$ is straightforward and we find the Laplace equation
\equn{laplace}
-\sigma \n. \left( \frac{\n h}{\sqrt{1+(\n h)^2}} \right) + \frac{\partial {\cal Q}}{\partial h} (h,Z) = 0
\nuqe
with condition (\ref{bc}). Note that this condition implies that {\it the equilibrium $h({\bf y})$ will be a functional of $\partial \Gamma$}. If we ignore gravity and the effective interface potential $V$, (\ref{laplace}) implies that the surface has constant mean curvature $\Delta p/2 \sigma$.

To consider variation with respect to the contact line we can without loss of generality specialize to the upper right-hand side quadrant of Fig. \ref{quad}. We assume that ${\bf y}$ on $\partial \Gamma$ can be written as ${\bf y} = (y_1, \gamma(y_1))$ for some function $\gamma(y_1)$ and for $0 \le y_1 \le Y$, where $Y$ is defined by $\gamma(Y) \equiv 0$ (see Fig.\ \ref{quad}). Thus, the function $\gamma(y_1)$ describes the shape of the contact line when projected onto the ${\bf y}$-plane.

For this region the free energy functional (\ref{f}) can be written as (in the Monge representation)
\eqan{intense}
F[h,Z] &=& \int_0^Y dy_1 \int_0^{\gamma(y_1)} dy_2 \Biggl \{ \sigma \sqrt{1+(\n h)^2} \nonu \\
& & - \Bigl [ \sigma_{w\alpha}({\bf y}) - \sigma_{w\beta}({\bf y}) \Bigr ] \sqrt{1+(\n Z)^2} + {\cal Q}(h;Z) \Biggr \} \nonu \\
& & + \int_0^Y dy_1 \lambda(y_1,\gamma(y_1)) \biggl[ 1+\gamma'(y_1)^2 \nonu \\
& & + \left( \frac{d}{dy_1} Z(y_1,\gamma(y_1)) \right)^2 \biggr]^\half 
\naqe
with the prime denoting differentiation with respect to the argument. The remainder of this part of the paper becomes by necessity quite mathematically intense and a casual reader may prefer to skip immediately to Sec.\ \ref{gye}.  

Functional minimization  of (\ref{intense}) leads to
\equn{xmas}
\frac{\delta F}{\delta \gamma(x)} = I_1 + I_2 + I_3
\nuqe
where
\eqan{midway}
I_1 &=&  \Biggl \{ \sigma \sqrt{1+(\n h)^2} - \Bigl [ \sigma_{w\alpha}({\bf y}) - \sigma_{w\beta}({\bf y}) \Bigr ] \nonu \\
& & \times \sqrt{1+(\n Z)^2} \left. + {\cal Q}(Z,Z) \Biggr \} \right|_{ {\bf y}=(x,\gamma(x))} \\
I_2 &=& \frac{\delta}{\delta \gamma(x)} \int_0^Y dy_1 \lambda(y_1,\gamma(y_1)) \biggl[ 1+\gamma'(y_1)^2 \nonu \\
& & + \left( \frac{d}{dy_1} Z(y_1,\gamma(y_1)) \right)^2 \biggr]^\half \nonu \\
\naqe
and
\equ
I_3 = \int_{\Gamma_\gamma} d{\bf y} \left\{ \sigma \frac{ \n h. \n \frac{\delta h}{\delta \gamma}}{\sqrt{1+(\n h)^2}} + \frac{\partial{\cal Q}}{\partial h} \frac{\delta h}{\delta \gamma} \right\} \label{mid2}
\uqe
The term $I_3$ arises from the implicit dependence of $h({\bf y})$ on $\gamma(x)$ and can be derived using the chain rule for functional derivatives.

We shall concentrate on simplifying the expression $I_2$ first. This becomes
\eqan{mess}
I_2 &=& \lambda \Bigl[(\gamma' Z_{y_1}-Z_{y_2})(Z_{y_1 y_1}+2\gamma' Z_{y_1 y_2} + \gamma^{'2} Z_{y_2 y_2}) \nonu \\
& &  - \gamma''(1+(\n Z)^2) \Bigr] \Bigl[ 1+ Z_{y_1}^2 + 2 \gamma' Z_{y_1} Z_{y_2} \nonu \\
& &  + \gamma^{'2}(1+Z_{y_2}^2) \Bigr]^{-\frac{3}{2}} + \Bigl[ \lambda_{y_2} (1+\gamma' Z_{y_1} Z_{y_2} + Z_{y_1}^2) \nonu \\
& & - \gamma' \lambda_{y_1} (1+Z_{y_1} Z_{y_2}/ \gamma' + Z_{y_2}^2) \Bigr] \Bigl[ 1+ Z_{y_1}^2 \nonu \\
& & + 2 \gamma' Z_{y_1} Z_{y_2} + \gamma^{'2}(1+Z_{y_2}^2) \Bigr]^{-\frac{1}{2}}
\naqe
where the notation $f_x = \frac{\partial f}{\partial x}$ is used and all functions are evaluated at ${\bf y}=(x,\gamma(x))$. Fortunately, (\ref{mess}) can be re-written as
\equn{a}
I_2 = \left. ( \n \lambda. \hat{\bf m} - \lambda C_g ) \sqrt{1+(\n Z)^2} \right|_{{\bf y} = (x,\gamma(x))}
\nuqe
where $\hat{\bf m}=\hat{\bf m}({\bf y})$ is the unit vector orthogonal to both the normal to the surface $z=Z({\bf y})$ and to the tangent vector to the curve $\partial \Gamma$. Here we use the convention that the $z$-component of $\n \lambda$ is zero. Furthermore, $C_g=C_g({\bf y})$ denotes the geodesic curvature of $\partial \Gamma$ at ${\bf y}$, i.e.\ the component of the `acceleration' vector of the curve in the direction $\hat{\bf m}$ \cite{book}.

Extending (\ref{mid2}) to the whole drop, one can use Gauss's theorem and write 
\eqan{here}
I_3 &=& \int_\Gamma d{\bf y} \left\{ \sigma \frac{ \n h. \n \frac{\delta h}{\delta \gamma}}{\sqrt{1+(\n h)^2}} + \frac{\partial{\cal Q}}{\partial h} \frac{\delta h}{\delta \gamma} \right\}  \nonu \\
&= & \sigma \oint_{\partial \Gamma} \frac{\delta h}{\delta \gamma} \left( \frac{h_{y_2} dy_1 - h_{y_1} dy_2}{\sqrt{1+(\n h)^2}} \right) \nonu \\
& & + \int_\Gamma d{\bf y} \left\{ -\sigma \n. \left( \frac{\n h}{\sqrt{1+(\n h)^2}} \right) + \frac{\partial {\cal Q}}{\partial h} \right\} \frac{\delta h}{\delta \gamma} \nonu \\
& = & \sigma \oint_{\partial \Gamma}  \frac{\delta h}{\delta \gamma} \left( \frac{h_{y_2} dy_1 - h_{y_1} dy_2}{\sqrt{1+(\n h)^2}} \right)
\naqe
by virtue of the Laplace equation (\ref{laplace}). Once again we now specialize to the upper right-hand quadrant. Here the boundary condition (\ref{bc}) can be expressed as
\equn{bc2}
h( y_1, \gamma(y_1) ) = Z(y_1, \gamma(y_1) )
\nuqe
Functionally differentiating (\ref{bc2}) with respect to $\gamma(x)$ leads to
\equn{help1}
\frac{\partial h}{\partial y_2} \delta(y_1-x) + \frac{\delta h}{\delta \gamma(x)} = \frac{\partial Z}{\partial y_2} \delta(y_1-x)
\nuqe
while differentiation with respect to $y_1$ gives
\equn{help2}
\frac{\partial h}{\partial y_1} + \frac{\partial h}{\partial y_2} \gamma'(y_1) = \frac{\partial Z}{\partial y_1} + \frac{\partial Z}{\partial y_2} \gamma'(y_1)
\nuqe
As we are on $\partial \Gamma$, $\gamma'(y_1) = \frac{dy_2}{dy_1}$ and so both (\ref{help1}) and (\ref{help2}) allow simplification of (\ref{here}). Using (\ref{help1}) to eliminate $\frac{\delta h}{\delta \gamma}$ and (\ref{help2}) to express everything in terms of $y_1$, the integral can be evaluated and (\ref{here}) becomes
\equn{b}
I_3 = \left. -\sigma \frac{\n h. \n (h-Z)}{\sqrt{1+(\n h)^2}} \right|_{{\bf y}=(x,\gamma(x))}
\nuqe

\section{The generalized Young equation}
\label{gye}
We are now in a position to interpret (\ref{xmas}). At equilibrium $\frac{\delta F}{\delta \gamma}$ must vanish and so using (\ref{a}) and (\ref{b}), (\ref{xmas}) implies
\eqan{young1} 
0 &=& \sigma \sqrt{1+(\n h)^2} -\sigma \frac{\n h. \n (h-Z)}{\sqrt{1+(\n h)^2}} + {\cal Q}(Z,Z) \nonu \\
& & - \sqrt{1+(\n Z)^2} \Bigl[ \sigma_{w\alpha}({\bf y}) - \sigma_{w\beta}({\bf y}) - \n \lambda. \hat{\bf m}({\bf y}) \nonu \\
& & + \lambda({\bf y}) C_g({\bf y}) \Bigr] 
\naqe
for ${\bf y} \in \partial \Gamma$. Recall that $C_g({\bf y})$ is the geodesic curvature of the contact line at the point $z = Z({\bf y})$ on the substrate surface and $\hat{\bf m}({\bf y})$ is the unit vector perpendicular to the surface normal and to the vector tangential to the triple line at ${\bf y}$. 

It follows from (\ref{P}) that ${\cal Q}(Z,Z) = 0$, which we use to write (\ref{young1}) as
\eqan{young2}
\lefteqn{\sigma \frac{1+\n h. \n Z}{\sqrt{ (1+(\n Z)^2)(1+(\n h)^2)}} =} & & \nonu \\
& &  \sigma_{w\alpha}({\bf y}) - \sigma_{w\beta}({\bf y}) - \n \lambda. \hat{\bf m}({\bf y}) + \lambda({\bf y}) C_g({\bf y})
\naqe
In fact, if we define the contact angle $\theta({\bf y})$ to be the angle between the normal (and hence the tangent) vectors to the surfaces $z=h({\bf y})$ and $z=Z({\bf y})$ at ${\bf y}$, then it is simple to show that
\equn{should}
\cos [\theta({\bf y})] = \frac{1+\n h. \n Z}{\sqrt{ (1+(\n Z)^2)(1+(\n h)^2)}}
\nuqe
Hence, (\ref{young2}) is the generalized Young equation for a sessile drop sitting on a rough, chemically heterogeneous surface $z=Z({\bf y})$,
\equn{Young}
\sigma_{w\alpha}({\bf y}) = \sigma_{w\beta}({\bf y}) + \sigma \cos [\theta({\bf y})] + \n \lambda. \hat{\bf m}({\bf y}) - \lambda({\bf y}) C_g({\bf y}) 
\nuqe
where ${\bf y}$ is at the three phase contact line. This is the main result of our paper. 

Particular cases of the generalized Young equation (\ref{Young}) have been obtained previously. If the line tension terms are ignored, one has 
\equ
\sigma_{w\alpha}({\bf y}) = \sigma_{w\beta}({\bf y}) + \sigma \cos \left[ \theta({\bf y}) \right]
\uqe
as derived by Lenz and Lipowsky \cite{peter}. 

The correction terms arising from the line tension have also been obtained previously for some special situations. First of all, their effect has been determined for planar and homogeneous substrates by Boruvka and Neumann \cite{bor} (and see also \cite{vess}). This case will be discussed in the following section. In addition, Rusanov \cite{rusanov1} has initiated study of an axially symmetric geometry for which the rotation axis is perpendicular to the ${\bf y}$-plane. In this case, the shape $Z$ of the substrate surface depends only on the distance $\rho=\sqrt{y_1^2 + y_2^2}$ from the rotation axis. Therefore, the wall tensions $\sigma_{wi} = \sigma_{wi}(\rho)$ and the line tension $\lambda = \lambda(\rho)$ are taken to depend only on $\rho$ and the problem becomes one dimensional. The contact line of the droplet term forms a circle of radius $\rho=R$, the unit vector $\hat{\bf m}$ has the $\rho$ component $m_\rho = 1/ \sqrt{1+Z'(R)^2}$, and the geodesic curvature is given by
\equ
C_g(R) = - \frac{1}{R \sqrt{1+Z'(R)^2}}
\uqe

If these expressions are inserted into the general equation (\ref{Young}), one obtains
\eqan{rus1}
\sigma_{w\alpha}(R) &=& \sigma_{w\beta}(R) + \sigma \cos [ \theta(R) ] + \left[ \frac{1}{R} \lambda(R) + \lambda'(R) \right] \nonu \\
& & \times \cos [ \phi(R)]
\naqe
where the slope angle $\phi$ satisfies
\equn{salpha0}
\cos [\phi(R)] = \frac{1}{\sqrt{1+Z'(R)^2}}
\nuqe
In order to understand the geometric meaning of this angle, consider the contour $z=Z(\rho)$ of the substrate surface within the $(\rho,z)$-plane, and construct the straight line which is tangential to $Z(\rho)$ at $\rho=R$. The slope angle $\phi(R)$ is the angle between this tangential line and the $\rho$ axis. The special form (\ref{rus1}) of the generalized Young equation (\ref{Young}) is equivalent to the equation derived by Rusanov \cite{rusanov1}.

Real substrates which are heterogeneous do not exhibit the axial symmetry, assumed in the derivation of (\ref{rus1}), and the shape of the contact line will not be circular. In order to incorporate the deviations of the line shape from a circle in a qualitative manner, Rusanov has proposed a generalization of (\ref{rus1}) which involves a heuristic line roughness factor \cite{rusanov2}.

Another effectively one dimensional geometry has been investigated by Marmur \cite{marmur,marmur2}. He has considered `cylindrical' interfaces which depend only on one surface coordinate, say $y_1$, and are translationally invariant in the $y_2$-direction. For this case, the contact line is perfectly straight and lies at $y_1= \pm Y$, for some constant $Y$ say, implying that the geodesic curvature $C_g(\pm Y) =0$. The unit vector $\hat{\bf m}$ has a $y_1$ component $m_{y_1} = 1/\sqrt{1+Z'(Y)^2}$ at $y_1=Y$ and so (\ref{Young}) becomes
\equ
\sigma_{w\alpha}(Y) = \sigma_{w\beta}(Y) + \sigma \cos[\theta(Y)] + \lambda'(Y) \cos[\phi(Y)]
\uqe
as stated in \cite{marmur2}.

\section{Planar and chemically homogeneous substrates}
To show that (\ref{Young}) includes the more familiar form of Young's equation is quite straightforward. We specialize to a smooth, homogeneous substrate, that is
\equ
\begin{array}{ccc} \sigma_{w\alpha}({\bf y}) = \sigma_{w\alpha} & ; & \sigma_{w\beta}({\bf y})=\sigma_{w\beta}  \\ \lambda({\bf y}) = \lambda & ; & Z({\bf y})=0
\end{array}
\uqe
For planar curves the geodesic curvature simply becomes the curvature, looking at (\ref{mess}) the first term tends to 
\equ
-\lambda C \equiv - \lambda \frac{\gamma''(x)}{[1+\gamma'(x)^2]^{\frac{3}{2}}}
\uqe
as $Z({\bf y})$ tends to a constant, while the vector $\hat{\bf m}({\bf y})$ in the same limit is the normal $\hat{\bf n}({\bf y})$ to the curve. Consequently, (\ref{Young}) is now
\equn{yg}
\sigma_{w\alpha} = \sigma_{w\beta} + \sigma \cos \theta  - \lambda C
\nuqe
This is Young's equation \cite{young} with the Boruvka and Neumann line tension term.

If corrections due to gravity and long-range intermolecular forces are ignored, the solution of the Laplace equation (\ref{laplace}) is a spherical cap with radius of curvature $R_\pi = 2 \sigma/\Delta p$. It is then easy to see that the curvature of the contact line itself satisfies
\begin{eqnarray}
C &=& -\frac{1}{R_\pi \sin \theta} \label{clc2} \\
&=& -\frac{\Delta p}{2 \sigma \sin \theta} \label{clc}
\end{eqnarray}
When this relation is inserted into (\ref{yg}), the contact angle $\theta$ is found to depend on the two dimensionless parameters $(\sigma_{w\alpha}-\sigma_{w\beta})/ \sigma$ and $\Delta p \lambda/ \sigma^2$. Note that, in the pressure ensemble considered here, the curvature $C$ of the contact line is fixed for a smooth, homogeneous substrate. 

Similarly, if one specifies a fixed volume $V$ of $\beta$ phase then again for {\it a smooth, homogeneous substrate} a relationship between $C$, $\theta$ and $V$ can be determined. Ignoring gravity and long-range forces, the droplet must take the form of a spherical cap. Consequently, it is easy to determine its volume and using (\ref{clc2}) relate this to the contact line curvature
\equn{cv}
C = -\left[ \frac{\pi(2-3 \cos \theta + \cos^3 \theta)}{3 V \sin^3 \theta} \right]^{1/3}
\nuqe
The contact angle once more depends on $(\sigma_{w\alpha}-\sigma_{w\beta})/ \sigma$ but now also on the new dimensionless quantity $\lambda / \sigma V^{1/3}$.

In principle, the line tension $\lambda$ may be determined via (\ref{yg}) if one knows the interfacial tensions $\sigma_{w\alpha}$, $\sigma_{w\beta}$ and $\sigma$ and if one measures the geometric parameters $\theta$ and $C$. The contact line curvature $C$ depends on $\theta$ both in the $\Delta p$- and in the $V$- ensembles. For the latter, it also depends on the droplet volume $V$, see (\ref{cv}).

Finally, thermally excited fluctuations away from the minimal free energy would in general change the contact line contour and so give additional corrections to $\theta$. Such effects are not included in this paper.

\section{Spatial Averaging}
\label{spatial}
Using the generalized Young equation (\ref{Young}) it is possible to provide a systematic derivation of the Cassie law (valid for flat but chemically heterogeneous substrates) and the Wenzel rule (applicable for purely rough surfaces),
\eqan{cas}
\cos \theta_{\rm eff} &=& \sum_{i} f_i \cos \theta_i \mbox{\hspace*{2cm} Cassie} \\
\cos \theta_{\rm eff} &=& r \cos \theta \mbox{\hspace*{2.8cm} Wenzel} \label{wenz}
\naqe
Here $\theta_{\rm eff}$ is an effective or average contact angle assumed on the heterogeneous substrate, $\theta_i$ is the angle taken on a simple planar surface composed entirely of surface component or chemical species $i$, $f_i$ is the fraction by area of the surface made up of $i$ and $r$ is the ratio of true to planar surface area. 

The starting point for a derivation of equations (\ref{cas}) and (\ref{wenz}) should be a definition of the effective or average angle $\theta_{\rm eff}$. However, in order to perform any average one must first establish choices for the following criteria:

\newcounter{d}
\begin{list}{(\roman{d})}{\usecounter{d}}
\item what statistical mechanical ensemble should be used and what are the different states in this ensemble?
\item what are the {\it a priori} probabilities or statistical weights assigned to these different states? Often, one makes an implicit assumption about equal {\it a priori} probabilities but this choice is not unique.
\item which quantities does one wish to average?
\end{list}

To answer question (i) we can choose to prescribe either the volume of the $\beta$ drop or else the pressure difference, $\Delta p$, across the $\alpha \beta$ interface. It is more convenient to use $\Delta p$ as the basic variable and so we opt to work in the corresponding pressure ensemble. Consequently, by ignoring gravitational and long-range intermolecular force corrections, we have droplet surfaces of constant mean curvature. For flat and homogeneous substrates, the latter are given by spherical caps and, from (\ref{clc}), have a fixed contact line curvature $C$. As discussed before, this is no longer true in the volume ensemble since $C$ depends on $V$, see (\ref{cv}). Therefore, if one wants to deduce the magnitude of the line tension $\lambda$ from observations of the droplet shape, it is important to specify the appropriate ensemble one is working in.

Having selected an ensemble we next need to decide what the different states are in this ensemble. Three different possibilities present themselves:
\newcounter{e}
\begin{list}{(\alph{e})}{\usecounter{e}}
\item One could investigate a drop at a certain position on the surface and move along its contact line. Aside from the flat and homogeneous case, the contact angle will vary and thus, one could study the average $\theta$ of an individual droplet. If a flat and homogeneous substrate is considered then the contact angle is constant satisfying (\ref{yg}). 
\item Different positions for the drop could be considered and the different states would then be determined by these positions. For the flat and homogeneous case this is the same as (a). However, if one places a drop at different positions on a heterogeneous substrate, the contact line contour and the contact angles at that contour will, in general, change if we require the state of the drop to be a local minimum of its free energy. The task of determining these states and then assigning {\it a priori} probabilities to them seems to be prohibitively difficult.
\item Instead of considering the states of the drop one could instead opt for those of the contact line. To start, a certain position ${\bf y}$ on the surface is selected and an average over all triple line contours which pass through this point {\it and} correspond to the states of the droplet which are local minima of the free energy is carried out. To be local minima these states must satisfy (\ref{laplace}), and so (if gravity and long-range intermolecular interactions can be ignored) be of constant mean curvature, as well as obeying the Young equation (\ref{Young}). A final further average over all positions ${\bf y}$ is then taken. 
\end{list}

In this section of the paper we choose option (c) above and proceed to average over different contact lines. To answer (ii) a Boltzmann weight is chosen for each contour, calculated via (\ref{f}), and again we point out that only those contours which are local minima of the free energy are to be considered.

From (\ref{Young}) one can see that the variation of the local contact angle $\theta({\bf y})$ is mainly governed by the change in the interfacial free energies $\sigma_{w\alpha}$ and $\sigma_{w\beta}$ (the line tension is typically $\approx 10^{-9} {\rm Jm}^{-1}$ \cite{drelich2}). However, these quantities determine the cosine of the contact angle and consequently summing over their position dependence should give an effective $\cos \theta_{\rm eff}$. Therefore, in response to (iii), we choose to average $\cos [ \theta({\bf y})]$ rather than $\theta({\bf y})$ itself. Arguing in terms of surface tensions, it is the component of the $\alpha \beta$ tension that is in the surface that is the one chosen to be averaged.

The Young equation (\ref{Young}) implies that the contact angle is dependent on position ${\bf y}$ and, as mentioned before, the local configuration taken by the triple line contour $\partial \Gamma$ at ${\bf y}$, that is 
\equ
\theta = \theta({\bf y};\partial \Gamma)
\uqe
To proceed, we define $\{ \partial \Gamma({\bf y}) \}$ to be the set of all contours of the drop which pass through ${\bf y}$ and are local minima of $(\ref{f})$. Then option (c) can be written mathematically as
\equn{bigav}
\cos \theta_{\rm eff} = \frac{ \int d{\bf y} \sum_{ \{ \partial \Gamma({\bf y}) \} } \cos \bigl( \theta({\bf y};\partial \Gamma) \bigr) {\rm e}^{-F \bigl[ h[\partial \Gamma],Z \bigr]/T}}{\int d{\bf y} \sum_{ \{ \partial \Gamma({\bf y}) \} } {\rm e}^{-F \bigl[ h[\partial \Gamma],Z \bigr]/T}}
\nuqe
where we use a Boltzmann weight function with units such that $k_B=1$ and write explicitly the functional dependence of the equilibrium $h=h[\partial \Gamma]$ on the contour configuration.

To make further progress several strong assumptions are needed. Firstly, the dependence of the contact angle on the shape of the droplet is ignored, i.e.\
\equn{assump1}
\cos [ \theta({\bf y};\partial \Gamma)] \approx \cos [\theta({\bf y})]
\nuqe
Secondly, the heterogeneities are taken to be such that the drop is not strongly confined to or repelled from any particular region. All non-homogeneities are `uniform' in this aspect. The probability that the contact line of the droplet passes through ${\bf y}$ is given by
\equ
\frac{\sum_{ \{ \partial \Gamma({\bf y}) \} } {\rm e}^{-F \bigl[ h[\partial \Gamma],Z \bigr]/T}}{ \int d{\bf y} \sum_{ \{ \partial \Gamma({\bf y}) \} } {\rm e}^{-F \bigl[ h[\partial \Gamma],Z \bigr]/T}} 
\uqe
and we assume that this is approximately independent of ${\bf y}$,
\equn{assump2}
\frac{\sum_{ \{ \partial \Gamma({\bf y}) \} } {\rm e}^{-F \bigl[ h[\partial \Gamma],Z \bigr]/T}}{ \int d{\bf y} \sum_{ \{ \partial \Gamma({\bf y}) \} } {\rm e}^{-F \bigl[ h[\partial \Gamma],Z \bigr]/T}} \approx \frac{1}{\int d{\bf y}} \equiv \frac{1}{A_\pi}
\nuqe
where $A_\pi$ represents the projected area of the substrate surface onto the ${\bf y}$-plane. No position is dramatically favoured over any other. Equation (\ref{assump2}) clearly drastically reduces the domain of validity of any results derived from (\ref{bigav}). Notice, that for a flat and homogeneous substrate (\ref{assump1}) and (\ref{assump2}) are satisfied identically.

Consequently, (\ref{bigav}) reduces and the effective angle now satisfies
\equn{theta}
\cos \theta_{\rm eff} \approx \frac{ \int d{\bf y} \cos [\theta({\bf y})]}{A_\pi}
\nuqe
which we use as a working definition.

\section{Planar but chemically heterogeneous surfaces}
We first specialise to smooth, chemically heterogeneous substrates. In the light of (\ref{assump1}), equation (\ref{Young}) is simply
\equn{damn}
\sigma_{w\alpha}({\bf y}) = \sigma_{w\beta}({\bf y}) + \sigma \cos [\theta({\bf y})] 
\nuqe
where all the line tensions terms are ignored. When this is integrated over the surface we are led to (recall (\ref{theta}))
\equn{cass1}
A_\pi \sigma \cos \theta_{\rm eff} = \int d{\bf y} [ \sigma_{w\alpha}({\bf y})-\sigma_{w\beta}({\bf y}) ]
\nuqe
where $A_\pi = \int d{\bf y}$. If our substrate has a fraction (by area) $f_i$ of substance $i$, then
\equ
\int d{\bf y} \left[ \sigma_{w\alpha}({\bf y})-\sigma_{w\beta}({\bf y}) \right]
 = A_\pi \sum_i f_i(\sigma_{i\alpha} - \sigma_{i \beta}) 
\uqe
where $\sigma_{i\alpha}$ is the wall-$\alpha$ interfacial free energy for a substrate chemically coated with material $i$, etc. Consequently, (\ref{cass1}) is
\equn{cass2}
\cos \theta_{\rm eff} = \sum_i f_i(\sigma_{i\alpha} - \sigma_{i \beta})/\sigma
\nuqe
which can then be interpreted in terms of more physical parameters using the Young equation (\ref{yg}). At this point, we would like to emphasize that (\ref{cass2}) already contains, at the level of the approximations used (see (\ref{assump1}) and (\ref{assump2})), {\it all} the physics associated with the heterogeneous substrate which is now completely characterised by the numbers $f_i$.  

Therefore, if, in the spirit of Cassie, we use 
\equn{yg2}
\sigma_{i\alpha}-\sigma_{i\beta} = \sigma \cos \theta_i - \lambda_i C_i
\nuqe
to re-express (\ref{cass2}) in the traditional form 
\equn{Cassie}
\cos \theta_{\rm eff} = \sum_i f_i \left( \cos \theta_i - \frac{\lambda_i C_i}{\sigma} \right)
\nuqe
then the values of $\theta_i$ and $C_i$ will, in general, have {\it no} relation to the original experimental set-up. The Young equation, (\ref{yg2}), is obeyed by all droplets adsorbed on a chemically homogeneous surface $i$. Therefore, provided $\theta_i$ and $C_i$ are measured for the {\it same} drop, it does not matter whether the pressure or volume ensemble is used, or what the pressure difference or volume of $\beta$ actually is. If one chooses to make the measurements on a heterogeneous substrate then it is vital that a point is selected for which the chemical composition of the surface can be unambiguously determined as $i$, and that {\it both} the contact angle and curvature of the contact line are measured {\it precisely at this point}.

 We feel that this step, from (\ref{cass2}) to (\ref{Cassie}), has not been highlighted in the literature, and, while it is certainly possible to determine $C_i$ and $\theta_i$ from a particular point of a complicated drop configuration present on a heterogeneous surface, there is no advantage in doing so. We emphasize that (\ref{cass2}) is the fundamental equation in which no ambiguities arise.

Note that the line tension terms in (\ref{Cassie}) are identical to those first included by Drelich and Miller \cite{drelich} by the use of thermodynamic arguments.

\section{Chemically homogeneous but rough substrates}
Wenzel's rule is taken to be valid for rough, chemically homogeneous substrates. However, from (\ref{Young}), when line tension terms are ignored, the contact angle taken on a rough surface (defined as the angle between the normal vectors to the surface and to the drop) is {\it identical} to that on a planar substrate. Consequently, we believe that (\ref{wenz}) in fact refers to the average of a {\it local} planar contact angle $\theta_\pi({\bf y})$, defined as
\equn{thetapi}
\cos [ \theta_\pi({\bf y})] = \left. \frac{1}{\sqrt{1+(\n h)^2}} \right|_{{\bf y} \in \partial \Gamma}
\nuqe
Thus, $\theta_\pi({\bf y})$ is the angle between the tangent to the drop's surface and a local horizontal plane given by ${\bf r} = ({\bf y}',Z({\bf y}))$ for all ${\bf y}'$.

\begin{figure}[h]
\begin{center}
\scalebox{0.5}{\includegraphics{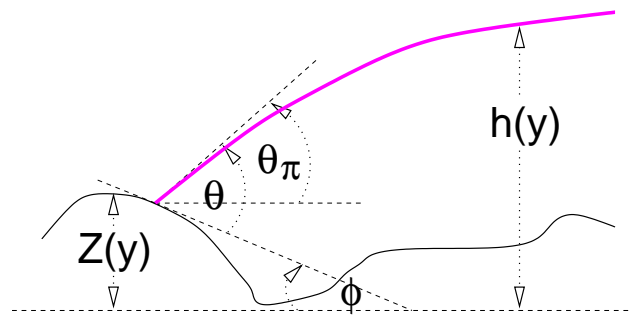}}
\end{center}
\caption{A cross section through a drop sitting on a one dimensional rough surface $z=Z(y)$. The true contact angle $\theta$, defined in (\ref{should}), the local angle, $\theta_\pi$, (\ref{thetapi}), and the slope angle $\phi$, (\ref{salpha}), are all clearly marked.}
\label{slope}
\end{figure}

Such an approach has a long history with, we believe, Shuttleworth and Bailey \cite{sb} first defining the observed contact angle to be the sum of the actual (or in their terminology intrinsic) contact angle and the slope angle, $\phi$ (see Fig.\ \ref{slope})
\equn{thetaob}
\theta_{\rm ob}({\bf y}) = \theta({\bf y}) + \phi({\bf y})
\nuqe
Here the slope angle is defined by
\equn{salpha}
\cos [ \phi({\bf y})] \equiv \frac{1}{\sqrt{1+[\n Z({\bf y})]^2}}
\nuqe
For a substrate surface shape $z=Z(x)$, which is dependent on only one surface coordinate $x$, it is straightforward to show that (\ref{thetapi}) and (\ref{thetaob}) are identical. For example, if the substrate surface is axially symmetric and its height is described by $z=Z(r)$, we obtain the slope angle $\phi$,  (\ref{salpha0}), which has the simple geometric interpretation explained in Sec.\ \ref{gye}. In a three dimensional space the decomposition (\ref{thetaob}) is not valid in general as there is no guarantee that $\phi$ and $\theta$ lie in the same plane. Our choice, $\theta_\pi$ or the slope angle of the $\alpha \beta$ interfacial surface, is more general and does not suffer from any such problems.

Proceeding with the definition (\ref{theta}), valid for $\theta({\bf y}) = \theta_\pi({\bf y})$, we write (\ref{Young}) as
\equn{wenie}
\sigma \cos[\theta_\pi({\bf y})] + \sigma \frac{\n h. \n Z}{\sqrt{1+(\n h)^2}} = [ \sigma_{w\alpha} - \sigma_{w\beta} ] \sqrt{1+(\n Z)^2}
\nuqe
where (\ref{assump1}) is again called upon to ignore line tension effects.
The surface area of the rough substrate, $A$, is
\equ
A = \int d{\bf y} \sqrt{1+(\n Z)^2}
\uqe
and therefore from (\ref{gauss}) and (\ref{laplace}), the integral of (\ref{wenie}) becomes
\equn{wenzelall}
A_\pi \sigma \cos \theta_{\rm eff} - \sigma \int d{\bf y} Z({\bf y}) \frac{\partial {\cal Q}}{\partial h}(Z,Z) = A(\sigma_{w\alpha}-\sigma_{w\beta}) 
\nuqe
or
\eqa
& \cos \theta_{\rm eff} -\frac{1}{A_\pi} \int d{\bf y} \left( \Delta p Z({\bf y}) +\Delta \rho g Z^2({\bf y}) \right) = & \nonu \\
& \frac{A}{A_\pi} \left( \cos \theta - \frac{\lambda C}{\sigma} \right) &
\aqe
using (\ref{P}) and (\ref{yg}), where $\theta$ and $C$ can again be measured for {\it any} drop located on a planar substrate. Notice that $Z({\bf y}) \goto 0$ as $y \goto \infty$ has been assumed. Denoting $r= \frac{A}{A_\pi}$ to be the ratio of non-planar to planar surface areas and utilizing the definition (\ref{plane})
we have
\equn{Wenzel}
\cos \theta_{\rm eff} = r \left( \cos \theta - \frac{\lambda C}{\sigma} \right) + \Delta \rho g \overline{Z^2}
\nuqe
where $\overline{Z^2}$ is the mean square height of the surface
\equ
\overline{Z^2} = \frac{\int d{\bf y} Z^2({\bf y})}{\int d{\bf y}}
\uqe
Equation (\ref{Wenzel}) is Wenzel's rule \cite{wenzel}, incorporating line tension effects and gravity for the first time. Here there is some disagreement between our line tension correction and that proposed by Drelich \cite{drelich2}. It is also interesting to note that the simple form of Wenzel given by (\ref{wenz}) was postulated a long time ago to become invalid under a gravitational field \cite{jd}. Finally, we again wish to emphasize that (\ref{wenzelall}) is independent of the properties of the drop being considered.

\section{Rough and chemically heterogeneous surfaces}
\label{both}

Continuing to use the definitions (\ref{theta}) and (\ref{thetapi}) for the effective contact angle, it is not too difficult to see that an additional law is possible for a sessile drop on a rough and chemically heterogeneous substrate. Equation (\ref{Young}), via (\ref{assump1}),
\eqa
& \sigma \cos[\theta_\pi({\bf y})] + \sigma \frac{\n h. \n Z}{\sqrt{1+(\n h)^2}} = & \nonu \\
& [ \sigma_{w\alpha}({\bf y}) - \sigma_{w\beta}({\bf y}) ] \sqrt{1+(\n Z)^2} &
\aqe
can be integrated over ${\bf y}$. Using (\ref{laplace}) one finds
\eqa
A_\pi \sigma \cos \theta_{\rm eff} &=& \sigma \int d{\bf y} Z({\bf y}) \frac{\partial {\cal Q}}{\partial h}(Z,Z) \nonu \\
& & + \int d{\bf y} \Bigl[ \sigma_{w\alpha}({\bf y})-\sigma_{w\beta}({\bf y}) \Bigr] \nonu \\
& & \times \sqrt{1+(\n Z)^2} 
\aqe
and so arrives at
\equn{CW}
\cos \theta_{\rm eff} = \sum_i r_i \left( \cos \theta_i - \frac{\lambda_i C_i}{\sigma} \right) + \Delta \rho g \overline{Z^2}
\nuqe
where $r_i$ is the ratio of the non-planar surface area covered with material $i$ to the total planar area.

\section{An alternative prescription}
\label{alternative}
It is somewhat unsatisfying that the traditional forms of the Cassie and Wenzel equations are independent of the shape of the drop and assume a `uniform' distribution of the heterogeneities as given by (\ref{assump2}). In this section we derive a new relation for the effective contact angle $\theta_{\rm eff}$ on a chemically heterogeneous substrate, which while not being quite as aesthetically pleasing, does not have these two drawbacks. Here we switch from the choice (c), described in  Sec.\ \ref{spatial}, of the states in our statistical mechanical ensemble to that of (a).

Let a drop of $\beta$ phase on a flat, chemically heterogeneous substrate equilibriate and take up the shape of the optimum contour, i.e.\ that which is a local minimum of the free energy. We then define the effective contact angle to be simply the average contact angle taken around this particular configuration of the three phase contact line. Writing $s$ for the arc length on $\partial \Gamma$ and the local contact angle then as $\theta({\bf y}) = \theta(s)$ we have 
\equ
\cos \theta_{\rm eff} = \frac{ \oint_{\partial \Gamma} ds \cos[ \theta(s)]}{\oint_{\partial \Gamma} ds}
\uqe
Integrating Young's equation in the form of (\ref{damn}) around the optimum contour gives
\eqa
\sigma \cos \theta_{\rm eff} \oint_{\partial \Gamma} ds &=& \oint_{\partial \Gamma} ds \Bigl[ \sigma_{w\alpha}(s) - \sigma_{w\beta}(s) \Bigr] \nonu \\
& & + \oint_{\partial \Gamma} ds \Bigl[ \lambda(s) C(s) - \n \lambda.\hat{\bf n}(s) \Bigr]
\aqe
Now looking again at the upper right-hand quadrant of Fig.\ \ref{quad}, the normal to the contact line is given by
\equ
\hat{\bf n}({\bf y}) = \frac{1}{\sqrt{1+\gamma'(y_1)^2}} \left( \begin{array}{c} -\gamma'(y_1) \\ 1 \end{array} \right)
\uqe
and so
\eqan{ohah}
\n . \hat{\bf n}({\bf y}) = - \frac{d}{dy_1} \left( \frac{\gamma'}{\sqrt{1+\gamma^{'2}}} \right) &=& - C(y_1,\gamma(y_1)) \nonu \\
& = & -C({\bf y})
\naqe
implying
\equ
\n . \hat{\bf n}(s) = -C(s)
\uqe
Hence, we have
\equn{yetmore}
\cos \theta_{\rm eff} = \sum_{i} \ell_i \left( \sigma_{i \alpha} - \sigma_{i \beta} \right) - \frac{\oint_{\partial \Gamma} ds \n. (\lambda(s) \hat{\bf n}(s))}{\oint_{\partial \Gamma} ds}
\nuqe
or
\equn{Cass2}
\cos \theta_{\rm eff} = \sum_{i} \ell_i \left( \cos \theta_i - \frac{\lambda_i C_i}{\sigma} \right) - \frac{\oint_{\partial \Gamma} ds \n. (\lambda(s) \hat{\bf n}(s))}{\oint_{\partial \Gamma} ds}
\nuqe
where $\ell_i$ is the fraction of the total {\it perimeter} of the drop which is on surface composed of material $i$ and $\lambda_i$ is the line tension for this surface. The angle $\theta_i$ and curvature $C_i$ are defined as before and can be measured for any drop on any flat surface $i$ providing both measurements are taken at the same point. Equation (\ref{Cass2}) can be written quite succinctly as
\equ
\cos \theta_{\rm eff} = \sum_{i} \ell_i \cos \theta_i 
\uqe
if line tension contributions are negligible.

From a theoretical point of view (\ref{Cass2}) is preferable to (\ref{Cassie}) as
\newcounter{c}
\begin{list}{(\roman{c})}{\usecounter{c} }
\item only one (the most likely) contour is considered
\item it is a local equation and it is only the surface in the immediate neighbourhood of the contact line that is needed. It would be obviously foolhardy to use (\ref{Cassie}) for a surface half of which is purely hydrophobic, for example, and the rest consisting of alternating hydrophobic and hydrophilic strips if the drop sits entirely in either of these two regions. Equation (\ref{Cass2}) requires no assumptions about the nature of the heterogenieties.
\item there is explicit dependence on the droplet shape due to the last integral term (remember that $\hat{\bf n}$ is the normal to the contact line)
\end{list}

Experimentally, during the last decade a variety of techniques have been developed in order to measure surface topographies on small scales. One pertinent example is provided by atomic or scanning force microscopy, which makes it possible to measure the precise shape of small droplets on nanometre scales and so, the exact location of the contact line, see e.g.\ \cite{herm}. Similar techniques can also be used in order to determine the chemical composition of the substrate surface. Consequently, one may obtain an estimate for the variation of the line tension $\lambda$ along the contact line $\partial \Gamma$. Therefore, it seems probable that the line integral in (\ref{Cass2}) can be determined experimentally from a careful analysis of scanning force micrographs. Hence, we can expect (\ref{Cass2}) to be of some practical use.

\section{Conclusion}
In this paper, by introducing a novel minimization scheme of the free energy, we have found the most general form of the Young equation. Equation (\ref{Young}) is valid for substrates both chemically and geometrically inhomogeneous, in the presence of long-range interactions and under the influence of gravity. Taking advantage of this new result, we have looked again at the phenomenological laws of Cassie and Wenzel and examined their statistical mechanical foundation. 

To recover (\ref{cas}) and (\ref{wenz}), two strong assumptions are needed: (i) the dependence of the contact angle on the shape of the drop has to be ignored and, (ii) all locations of the droplet are taken to be equally likely. These two assumptions lead to the estimate (\ref{theta}) for the effective contact angle.

A new insight, that our approach has brought to light, is the interpretation of the contact angle $\theta_i$ and triple line curvature $C_i$ in the Cassie, (\ref{Cassie}), and in the Wenzel, (\ref{Wenzel}), equations. Recall that here the index $i$ distinguishes between the different compounds or chemical species in the substrate surface. When using these results it is vitally important to realize that $\theta_i$ and $C_i$ are present in the exact combination described by the planar Young's equation, (\ref{yg}), i.e.\ for each substrate component $i$, $\sigma \cos \theta_i - \lambda_i C_i$ always appears, which is identically equal to $ \sigma_{i\alpha}-\sigma_{i\beta}$. Therefore, when we wish to understand (\ref{Cassie}) and (\ref{Wenzel}) on a practical level, the effective contact angle is determined by the ratios $f_i$ or $r$ and by the value of $\sigma_{i\alpha}-\sigma_{i\beta}$ for the surface $i$ under investigation. Due to the universality of Young's equation, (\ref{yg}), the actual $\theta_i$ and $C_i$ used are irrelevant, in the sense that they need bear no relation to the droplet shape observed in our current experiment. All that is required, is that they are measured at the same position in space for any drop of $\beta$ in $\alpha$ phase, providing it lies, at that point, on a surface domain composed of the substrate compound $i$. Young's equation then guarantees that the measured values of $\theta_i$ and $C_i$, when used in the Cassie or Wenzel laws, will simply combine to give $\sigma_{i\alpha}-\sigma_{i\beta}$ as required. To reiterate, due to the particular way they occur in (\ref{Cassie}) and (\ref{Wenzel}), $\theta_i$ and $C_i$ have, in general, no relation to the actual shape of the droplet on a heterogeneous substrate.

Within this remit, the relation (\ref{CW}) is proposed which relates the effective contact angle on a substrate with both geometric and chemical heterogeneities to those on simple planar surfaces. We also find the new relation (\ref{Cass2}) for the effective contact angle on a chemically heterogeneous substrate, which is true generally and requires none of the above assumptions. Consequently, the influence of the surface on the shape of the wetting droplet {\it is} taken into account. As discussed in the previous section, the line integrals which appear in the new relation (\ref{Cass2}) for the effective contact angle can be estimated from scanning force microscopy data of the droplet shape. Such an alternative approach does not recover a local form of Wenzel's rule, possibly because this rule does not involve an average of the true contact angle but rather of a local planar one.

\acknowledgments
We are happy to thank Peter Lenz for (many) useful discussions and several anonymous referees for providing interesting comments.


\begin{references}
\bibitem[*]{email} Email swain@mpikg-teltow.mpg.de
\bibitem[\dag]{email2} Email lipowsky@mpikg-teltow.mpg.de
\bibitem{rev} G.\ H.\ Findenegg and S.\ Herminghaus, Curr.\ Opin.\ Colloid Interface Sci.\ {\bf 2}, 301 (1997)

E.\ M.\ Blokhuis and B.\ Widom, Curr.\ Opin.\ Colloid Interface Sci.\ {\bf 1}, 424 (1996)

G.\ Forgacs, R.\ Lipowsky and T.\ M.\ Nieuwenhuizen, in {\it Phase  Transitions and Critical Phenomena Vol.\ 14}, edited by C.\ Domb and J.\ L.\ Lebowitz (Academic Press, London), 1991.
\bibitem{wenzel} R.\ N.\ Wenzel, J.\ Phys.\ Colloid Chem.\ {\bf 53}, 1466 (1949) and Indust.\ Eng.\ Chem.\ {\bf 28}, 988 (1936).
\bibitem{cassie} A.\ B.\ D.\ Cassie, Discuss.\ Faraday Soc.\ {\bf 3}, 11 (1948).
\bibitem{borgs} C.\ Borgs, J.\ de Coninck, R.\ Koteck\'{y} and M.\ Zinque, Phys.\ Rev.\ Lett.\ {\bf 74}, 2292 (1995).
\bibitem{us} A.\ O.\ Parry, P.\ S.\ Swain and J.\ A.\ Fox, J.\ Phys.\ Condens.:\ Matter {\bf 8}, L659 (1996)

P.\ S.\ Swain and A.\ O.\ Parry, J.\ Phys.\ A {\bf 30}, 4597 (1997)

P.\ S.\ Swain and A.\ O.\ Parry, Eur.\ Phys.\ J.\ B, in press (1998).

\bibitem{urban} D.\ Urban, K.\ Topolski and J.\ de Coninck, Phys.\ Rev.\ Lett.\ {\bf 76}, 4388 (1996).
\bibitem{tosh} B.\ V.\ Toshev, D.\ Platikanov and A.\ Scheludko, Langmuir {\bf 4}, 489 (1988).
\bibitem{drelich} J.\ Drelich and J.\ D.\ Miller, Langmuir {\bf 9}, 619 (1993).
\bibitem{drelich2} J.\ Drelich, Colloids Surfaces A{\bf 116}, 43 (1996). 
\bibitem{young} T.\ Young, Philos.\ Trans.\ R.\ Soc.\ London {\bf 95}, 65 (1805).
\bibitem{rowlinson} J.\ S.\ Rowlinson and B.\ Widom, {\em Molecular Theory of Capillarity}, (Clarendon Press, Oxford), 1982.
\bibitem{peter} P.\ Lenz and R.\ Lipowsky, Phys.\ Rev.\ Lett.\ {\bf 80}, 1920 (1998).
\bibitem{vdw} R.\ Lipowsky, Phys.\ Rev.\ B {\bf 32}, 1731 (1985).
\bibitem{book} J.\ A.\ Thorpe, {\it Elementary Topics in Differential Geometry} (Springer-Verlag, New York), 1979.
\bibitem{bor} L.\ Boruvka and A.\ W.\ Neumann, J.\ Phys.\ Chem.\ {\bf 66}, 5464 (1977).
\bibitem{vess} V.\ S.\ Vesselovsky and V.\ N.\ Pertzov, Zh.\ Fiz.\ Khim.\ {\bf 8}, 245 (1936).
\bibitem{rusanov1} A.\ I.\ Rusanov, Colloid J.\ USSR (Engl.\ Transl.) {\bf 39}, 618 (1977).
\bibitem{rusanov2} A.\ I.\ Rusanov, Mendeleev Commun.\ {\bf 1}, 30 (1996).
\bibitem{marmur} A.\ Marmur, Langmuir {\bf 12}, 5704 (1996).
\bibitem{marmur2} A.\ Marmur, Colloids Surfaces A {\bf 136}, 81 (1998).
\bibitem{sb} R.\ Shuttleworth and G.\ L.\ J.\ Bailey, Discuss.\ Faraday Soc.\ {\bf 3}, 16 (1948).
\bibitem{jd} R.\ E.\ Johnson and R.\ H.\ Dettre, Advances in Chemistry Series {\bf 43}, 112 (1964).
\bibitem{herm} S.\ Herminghaus, A.\ Fery and D.\ Reim, Ultramicroscopy {\bf 69}, 211 (1997).
\end{references}
\end{document}